\documentclass{article}

\usepackage{epsfig}
\usepackage{amsmath}   
\usepackage{amsfonts}
\usepackage{amssymb}
\usepackage{graphicx}  
\usepackage{verbatim}
\usepackage{color}   
\usepackage{subfigure}
\usepackage{hyperref}
\usepackage{authblk} 
\usepackage{newtxtext,newtxmath} 

\newcommand{\citen}[1]{\cite{#1}}

\makeindex

\newcommand{\wt}{\widetilde}

\begin{document}

\title{Normal forms, universal scaling functions, and extending the 
validity of the RG \label{SethnaChapter}\\
{\it Contribution to the book commemorating Michael Fisher's contributions
to condensed-matter physics and critical phenomena.
}}

\author{James P. Sethna}
\affil{LASSP, Cornell University, Ithaca, NY 14853, USA}

\author{David Hathcock}
\affil{IBM T. J. Watson Research Center, Yorktown Heights, NY 10598}

\author{Jaron Kent-Dobias}
\affil{INFN, Unità di Roma 1, 00185 Rome, Italy}

\author{Archishman Raju}
\affil{Simons Centre for the Study of Living Machines, National Centre for Biological Sciences, Tata Institute of Fundamental Research, Bengaluru 560065, India}

\maketitle

\begin{abstract}
Our community has a deep and sophisticated understanding of phase transitions
and their universal scaling functions.
We outline and advocate an ambitious program to use this understanding
as an anchor for describing the surrounding phases.  We explain how to use normal form theory to write universal
scaling functions in systems where the renormalization-group flows cannot
be linearized. We use the 2d Ising model to demonstrate how to calculate
high-precision implementations of universal scaling functions, and how to
extend them into a complete description of the surrounding phases. We 
discuss prospects and challenges involved into extending these early successes
to the many other systems where the RG has successfully described emergent
scale invariance, making them invaluable tools for engineers, biologists, and social scientists studying complex systems.
\end{abstract}

\section{Introduction}
\label{sec:SethnaIntro}

Half a century ago, Michael Fisher, together with Wilson and Kadanoff,
introduced the renormalization group to analyze systems with emergent,
fractal scale
invariance. For five decades, physicists have applied these techniques
to equilibrium phase transitions, avalanche models, glasses and
disordered systems, the onset of chaos, plastic flow in crystals,
surface morphologies, etc. But these tools have not made a substantial
impact on engineering and biology. We believe it is our duty to make these
tools accessible to the broader science community.

$\bullet$
We need to provide them with tools that allow them to describe
not only the critical point, but properties of systems that 
exhibit incipient scale-invariant fluctuations yet are
far from the critical region. These demand that we understand
{\em corrections to scaling}, which become more important farther from
the critical point. In section~\ref{sec:AnalyticCorrections}, we build
on Michael's work on analytic corrections to scaling with
Aharony~\cite{AharonyF80,AharonyF83} and his work on the complex analytic
properties of the Onsager solution~\cite{FisherCircleOfZeros}
to extend our understanding of the 2d Ising critical point to
a full description of the ferromagnetic and paramagnetic
phases~\cite{HathcockSnn}.

$\bullet$
We need not only universal power laws, but a complete
description of all behavior of the material. These demand convenient access to 
accurate {\em universal scaling functions} that govern behavior involving more
than two quantities at a time. Michael taught us about these powerful 
tools in his pedagogical reviews and
lectures.
So, magnetization as a function of
temperature goes as $M \sim t^\beta$, but magnetization as a function of
temperature and external field goes as 
$M \sim t^\beta \mathcal{M}(h/t^{\beta \delta})$, where
$\mathcal{M}(X)$ is an universal function, in principle predicted from the
renormalization group. 
In section~\ref{sec:NormalForms}, we discuss
the correct way~\cite{RajuCHKLRS19} of writing universal scaling
functions for systems where logarithms and exponentials
are found in addition to power laws (e.g., in the upper and lower critical dimensions).
In section~\ref{sec:2dIsingInField}, we present
a user-friendly solution~\cite{KentDobiasSnn} for $\mathcal{M}(X)$
in the 2d Ising model,
with a systematic expansion that captures the correct singular behaviors
with seven-digit accuracy. 

$\bullet$ 
We need rapidly converging methods that can connect our deep understanding
of the singularities in phases and in universal scaling functions to 
quantitative predictions which make the best use of limited information
away from the critical points. 
In modern numerical analysis, one can integrate or approximate analytic
functions of one variable with exponential
accuracy~\cite[secs.~4.6,5.8.1]{PressTVFNR} so long as one knows the
singularities at the endpoints. Chebyshev, Gauss, and Romberg methods have 
superseded Simpson's rule and its generalizations for approximating and integrating analytic 
functions, and can be adapted to capture known singularities. We know
that all properties are analytic in phases, and our mission for five
decades has been to understand the singularities between phases. Can
we explain the phases using the critical points?
The exponential convergence in
sections~\ref{sec:AnalyticCorrections} and~\ref{sec:2dIsingInField} 
demonstrate that the normal form theory can do so.

Focusing on the 2d Ising model allows us to show proof of principle, 
but can we aspire to similar progress for other, less well studied
critical points? We suggest this as a key task for the next stage of 
research in critical phenomena and emergent scale invariance. In
section~\ref{sec:NPFRG}, we discuss
progress by those using NPFRG (non-perturbative
functional renormalization-group) methods~\cite{NPFRG21},
which have found broad application not only in equilibrium thermodynamic
systems, but in avalanche models, quantum systems, turbulence, \dots
They explicitly coarse-grain and renormalize a system systems in fixed
dimension, implicitly calculating the universal scaling functions
and the system-specific behavior far from criticality. They have mostly
been used to extract high-precision critical exponents, amplitude ratios,
and proofs of success. If we can organize these calculations into
user-friendly universal scaling functions, we may provide experimentalists,
simulators, and theorists in a variety of fields with tools to describe
matter beyond systems tuned to a single point on the phase diagram.

In the final section~\ref{sec:ProspectsForFuture}, we embark into deep 
issues in the renormalization group and the prospects and challenges they
provide to our mission to make the theory of critical phenomena an organizing
principle for much of science.

\section{Normal form theory, RG flows, and logs}
\label{sec:NormalForms}

The renormalization group takes an enormous leap of abstraction -- studying
emergent scale invariance as a flow under coarse-graining in the space of
all possible systems. This reduces the problem to the study of fixed points
for differential equations in an infinite-dimensional space (e.g., free energy
$f$, temperature $t=(T-Tc)$, field $h$, other parameters $u$, \dots). Near the
critical point for the Ising model, coarse-graining
and rescaling the free energy $f$ by a factor $e^\ell$ is described by the
differential equations
\begin{equation}
\label{eq:NLRGFlow}
\begin{aligned}
df/d\ell &= d f + A t^2 + \mathrm{other~nonlinear\dots}\\
dt/d\ell &= \lambda_t t + \mathrm{nonlinear\dots}\\
dh/d\ell &= \lambda_h h + \mathrm{nonlinear\dots}\\
du/d\ell &= \lambda_u u + \mathrm{nonlinear\dots}
\end{aligned}
\end{equation}
The first step in most treatments of these RG flows is to linearize the
flow near the fixed point. Positive eigenvalues $\lambda_t$ and $\lambda_h$
correspond to relevant operators like field $h$ and temperature $t$; negative 
eigenvalues $\lambda_u$ correspond to `irrelevant' perturbations
like $u$ that provide {\em singular corrections to scaling}
that are subdominant near the critical point. 
Using this linearization, this treatment then argues for universal power laws
for things like the correlation length $\xi(t) \sim t^\nu = t^{1/\lambda_t}$
at $h=0$, and universal power laws times universal functions
\begin{equation}
\label{eq:Fscaling}
f(t,h,u) \sim t^{d \nu} \mathcal{F}(h/t^{\beta \delta}, u t^\omega) 
= t^{d/\lambda_t} \mathcal{F}(h/t^{\lambda_h/\lambda_t},u t^{-\lambda_u/\lambda_t}),~\mathrm{and}
\end{equation}
\begin{equation}
\label{eq:Mscaling}
M(t,h,u) \sim t^\beta \mathcal{M}(h/t^{\beta \delta}, u t^\omega) 
= t^{(d-\lambda_h)/\lambda_t} \mathcal{M}(h/t^{\lambda_h/\lambda_t},u t^{-\lambda_u/\lambda_t})
\end{equation}
for things like the field-dependent free energy and magnetization, that describe
relations between more than two quantities.

This linearization is not useful in many cases (e.g., the 1d, 2d, and 4d Ising
models and all models in their upper and lower critical dimensions). How
to systematically formulate universal scaling functions in these cases
has hitherto been mysterious.
In this section, we describe the use of normal form methods
from dynamical systems theory by Raju et al.~[\citen{RajuCHKLRS19}] 
to understand when
this linearization is possible, and how to modify the invariant arguments to
the universal scaling functions functions when it is not.

Wegner and co-workers~\cite{Wegner72,WegnerR73} in the early days justified this
linearization by changing variables to {\em nonlinear scaling fields}
which transform linearly under the renormalization group. 
Cardy~\cite{Cardy96} denotes these new variables
$u_t$ and $u_h$, but we shall use tildes $\wt{t}$ and $\wt{h}$. By choosing
a suitable Taylor expansion of the change of variables,
\begin{equation}
\label{eq:COV}
\begin{aligned} 
t(\wt{t},\wt{h},\wt{u}\dots) &= \wt{t} + a_{tu} \wt{t} \wt{u} + a_{th^2} \wt{t} \wt{h}^2 + \dots\\
h(\wt{t},\wt{h},\wt{u}\dots) &= \wt{h} + b_{hu} \wt{h} \wt{u} + b_{ht} \wt{h} \wt{t} + \dots\\
\dots
\end{aligned}
\end{equation}
the equations simplify to 
$d\wt{f}/d\ell = d \wt{f}$, $d\wt{t}/d\ell = \lambda_t \wt{t}$, $d\wt{u}/d\ell =\lambda_u \wt{u}$, etc. 
Aharony and Fisher~\cite{AharonyF80,AharonyF83} ten years later noted that
this change of variables leads to what we call {\em analytic corrections to
scaling}, again subdominant near the critical point. The analytic corrections
to scaling have power laws that involve integers and combinations of
existing critical exponents $\beta$, $\nu$, $\delta$, and include an analytic
background to the free energy; these corrections can be written in terms of
derivatives of the universal scaling function. The aforementioned singular corrections to scaling introduce new critical exponents, and become independent
variables in the universal scaling functions.

Dynamical systems theory~\cite{Wiggins03,RajuCHKLRS19} tells us that
linearizing the
flow can only be done for what are called hyperbolic fixed points. The
change of variables $f,t,h,u \to \wt{f},\wt{t},\wt{h},\wt{u}$ is calculated
one polynomial order at a time, and its radius of convergence can be a subtle
mathematical issue~\cite{Guckenheimer}. In
section~\ref{sec:AnalyticCorrections}, we shall take on the ambitious task of 
attempting to describe the entire surrounding paramagnetic and ferromagnetic
phases for the 2d Ising model. There we shall see that convergence is tricky,
but a good choice of coordinates can yield a radius of convergence that 
appears to converge precisely in the range from zero to infinite temperature.

Even for the two-dimensional Ising model, the RG fixed point cannot be 
linearized. The specific heat has a logarithmic singularity: often 
deemed $\alpha = 0 (\log)$, but incompatible with a linearized flow. 
This arises because in 2d no polynomial change can remove
the $A t^2$ term
in the flow of $f$ in Eq.~\ref{eq:NLRGFlow} (although rescaling the relative
magnitudes of $f$ and $t$ can change the value of $A$ to one). Somewhat
messy algebra can confirm that this is due to a integer {\em resonance}
between the two linear eigenvalues
$\lambda_f = d = 2 = 2/\nu = 2*\lambda_t$ for the 2d Ising model.
The simplest 
normal form is thus
$d\wt{f}/d\ell = 2 \wt{f} - \wt{t}^2$, $d\wt{t}/d\ell = \wt{t}$, etc. This
results in a singularity in the free energy of the form $t^2 \log(t^2)$ which
will play an important role in sections~\ref{sec:AnalyticCorrections} 
and~\ref{sec:2dIsingInField}.

For the 4d Ising model, the leading irrelevant operator $u$ becomes marginal,
with a zero eigenvalue $\lambda_u$. There it has always been clear that one
cannot linearize the RG flow. In the dynamical systems nomenclature, the
flow undergoes a transcritical bifurcation in $d=4$ (together with the
same resonance $d = 2/\nu$ seen in 2d Ising above). Our
analysis~\cite{RajuCHKLRS19} shows that the normal form for the RG flows 
is%
  \footnote{\label{foot:NonlinearTermHZero}
  The nonlinear term proportional to $\wt{u}\wt{h}$ in the equation
  for $d{\wt{h}}/d{\ell}$ needed in the general normal form is known to be
  zero for the 4d Ising model. Raju~\cite{RajuPhD} has shown this is because there is a redundant variable proportional to the magnetization cubed in the renormalization group (see sec.~\ref{sec:ProspectsForFuture}).}
\begin{equation}
  \label{eq:FlowEquationIsing4D}
  d{\wt{f}}/d{\ell} = 4 \wt{f} - \wt{t}^2, ~~~ 
  d{\wt{t}}/d{\ell} = 2\wt{t} - A \wt{u} \wt{t}, ~~~
  d{\wt{h}}/d{\ell} = 3\wt{h}, ~~~
  d{\wt{u}}/d{\ell}  = -\wt{u}^2 + D \wt{u}^3.
\end{equation}
The universal scaling of the free energy in a system of length $L$
does not take the usual scaling form
$f(t,h,u,L) = L^{-d} \mathcal{F}(X,Y,Z) + f_a(t,h,u,L)$
with $X = \wt{t} L^2$, $Y = \wt{h} L^3$, and $Z=\wt{u}/L^{\omega/\nu}$.
First, we have quite unusual scaling variables
\begin{equation}
  \label{eq:InvariantCombinations4D}
  X = \wt{t} L^2 \left(W(y L^{1/D})/(1/(D \wt{u})-1)\right)^{-A},
  Y = \wt{h} L^{3}, ~~\mathrm{and}~~
  Z = y L^{1/D}
\end{equation}
Second, the free energy has a more complex form
\begin{equation}
f(t,h,u,L) = L^{-d} f(X,Y,Z) 
	- W(Z)^{-A} \left(\frac{W(Z)^{-A}}{1-A}-\frac{1}{A}\right)
	+ f_a(t,h,u,L).
\end{equation}
Here $W(x)$ is the Lambert $W$ or product log function, $\wt{u}$ is the marginal
quartic term in the Landau free energy, $y$ is a messy known
function of $\wt{u}$, $A$ and $D$ are the
amplitudes of nonlinear terms in the renormalization group flow
(Eq.~\ref{eq:FlowEquationIsing4D}), and
$f_a$ is a non-singular, analytic function near the critical point.

Eq.~\ref{eq:InvariantCombinations4D} captures the complete,
correct singularity for the 4d Ising model; the traditional log and
log-log corrections in the specific heat and the susceptibility arise from
expansions of the Lambert $W$ function for large arguments. At bifurcations
like $d=4$ and resonances as at $d=2$, normal form theory dictates the
nature of the singularity at the critical point. 

It {\em also} tells
us that the free energy in the surrounding phase can be found by changing
variables. So, for example, we know that the liquid-gas critical point
at high pressures and temperatures is in the 3d Ising universality class. 
Hence the liquid and gas phase properties should be given by
\begin{equation}
f(T,P) = \wt{t}(T,P)^{3 \nu} 
\mathcal{F}_\mathrm{3dIsing}(\wt{h}(T,P)/\wt{t}(T,P)^{\beta \delta},
				\wt{u}(T,P) t^\omega, \dots) + f_a(T,P),
\end{equation}
where $f_a$, $\wt{t}$, $\wt{h}$, and $\wt{u}$ are analytic in temperature and 
pressure near the critical point.
As usual, from the free energy
(and a corresponding scaling form for the correlation function) one can
derive all of the equilibrium and linear-response properties of the
resulting phases. We shall implement a change of coordinates like this 
for the 2d Ising model in section~\ref{sec:AnalyticCorrections}.

\begin{figure}
\centerline{\includegraphics[width=0.75\textwidth]{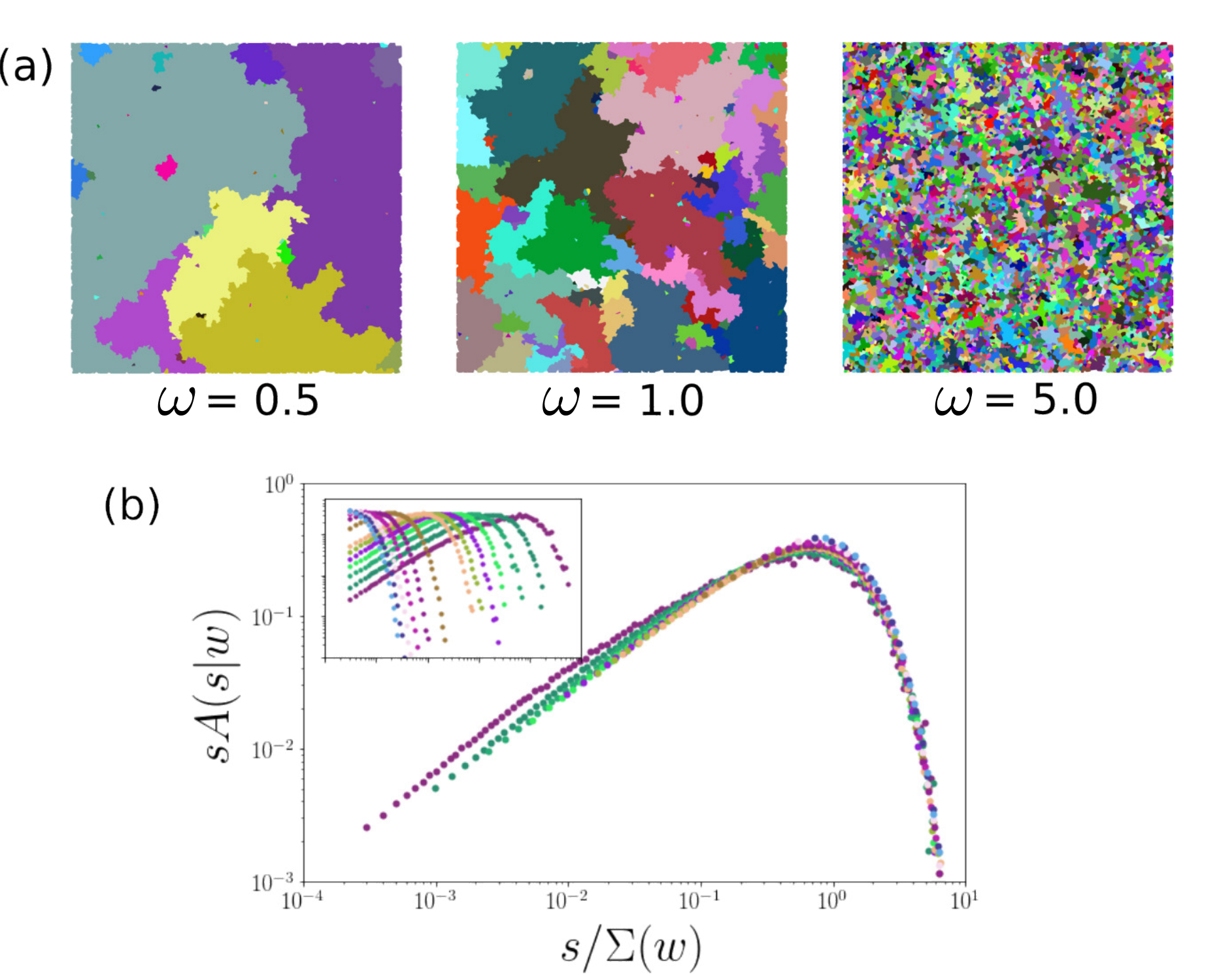}}
   \caption{{\bf Avalanche sizes} for the 2d $T=0$ random-field Ising model
   (from~[\citen{HaydenRS19}]).
   (a)~{\bf Avalanches}
   for disorders
   $w = 0.5$, $1.0$, and $5.0$; each color is a separate avalanche.
   (b)~{\bf Scaling collapse} of the area-weighted avalanche
   size distribution. 
   Note the factor of ten range in disorder $w$, and the
   factor of 2000 range in typical avalanche size $\Sigma$.
   Here the invariant scaling combination
   $\Sigma(w) = \Sigma_s (B+1/w)^{-B d_f+C}\exp(d_f/w)$ is not a ratio
   of power laws, but is derived directly from the nonlinear normal
   form of the renormalization-group equations in the
   lower critical dimension (as in Eq.~\ref{eq:FlowEquationIsing4D}).
   The area-weighted avalanche size distribution is thus
   $s A(s|w) = (s/\Sigma(w))^x \mathcal{A}((s/\Sigma(w))^y)$, with
   $\mathcal{A}$ a universal scaling function.}
   \label{fig:AvalancheSizeCollapse}
\end{figure}

We highlight the use of our normal-form methods~\cite{RajuCHKLRS19}
to solve a twenty-year-old
puzzle -- the unusual scaling of avalanches in the non-equilibrium
2d random-field
Ising model~\cite{HaydenRS19} (see Fig.~\ref{fig:AvalancheSizeCollapse}).
Over 25 years ago, we used the random-field Ising model to study avalanches. 
We could understand the scaling in 3, 4, 5, 7, and~9
dimensions~\cite{PerkovicDS99}, but two dimensions made no sense. Changing
the arguments from ratios of power laws to the invariant functions determined
from normal form theory was one part of the puzzle.%
  \footnote{Using random lattices to suppress faceting was the other obstacle.}

The distribution of avalanche sizes in the random-field Ising model is
cut off at size $\Sigma(w)$ depending on the disorder $w$. 
In three dimensions and higher it takes
the traditional form $\Sigma(w) = w^{-d_f \nu}$. In two dimensions, normal form
theory predicts that $\Sigma = (B+1/w)^{-B d_f + C} \exp(d_f/w)$, with
$d_f$ a universal critical exponent, and $B$ and $C$ being universal
constants associated with irremovable nonlinear terms in the
renormalization-group flow. Using this, Fig.~\ref{fig:AvalancheSizeCollapse}
does an excellent job explaining the behavior.

\section{Changing coordinates to describe phases: Matching Onsager}
\label{sec:AnalyticCorrections}

As foreshadowed in section~\ref{sec:NormalForms}, normal form theory tells us how to transform universal scaling forms to describe the entire phases surrounding a critical point. This section summarizes our recent results~\cite{HathcockSnn}, which implements this for the 2d Ising model in zero field, where Onsager's exact solution~\cite{onsager1944Exact2D} for the free energy enables quantitative validation of the results.

Given the asymptotic scaling form for the free energy (or any other quantity) in normal form coordinates, $\wt f(\wt t, \wt h, \wt u)$, the free energy as a function of physical temperature, field, and irrelevant perturbations is $f(t,h,u) = \wt f(\wt t(t,h,u), \wt h(t,h,u), \wt u(t,h,u)) + f_a(t,h,u)$. The functions $\wt t(t,h,u)$ and $\wt h(t,h,u)$, are given by precisely the analytic change of variables that inverts Eq.~\ref{eq:COV}, mapping the nonlinear scaling variables back to their physical counterparts. Additionally, we must also add terms $f_a(t,h,u)$ accounting for the analytic background of the free energy. Note that the change of coordinates are linear at lowest order: for example, $\wt t \sim t$ and $\wt h \sim h$. Therefore, $f(t,h,u) \sim \wt f(t,h,u)$ near the critical point; in other words, the normal form free energy $\wt f$ is the asymptotic scaling form near the critical point, usually computed using RG and related techniques.

We have already heard that the 2d Ising model in zero field has a
nonlinear normal form due to the resonance between the free energy
and temperature, 
$d\wt f/d\ell = 2 \wt f - \wt t^2$, $d \wt t/d\ell = \wt t$,
 and that these flow equations give rise to a logarithmic singularity,
$\wt f = -\wt t^2 \log(\wt t^2)$. Following the procedure outlined above,
the free energy as a function of temperature is simply
\begin{equation}\label{eq:2dfreeEnergy}
f(t) = -\wt t(t)^2 \log(\wt t(t)^2) + f_a(t). 
\end{equation}
This result is consistent with Onsager's exact solution, which is known to have the form $a(t) \log t^2 + b(t)$ for some analytic functions $a$ and $b$. Comparing these expressions for the free energy, we need to find the coordinate change $\wt t(t)$ and analytic background $f_a(t)$ by solving
\begin{equation}\label{eq:2dCOV}
a(t) = -\wt t(t)^2,  \quad \quad  b(t) = \wt t(t)^2 \log( \wt t(t)^2/t^2) + f_a(t).
\end{equation}
Note that because $\wt{t}(t)$ is linear to lowest order in t, the term $\log\left(\wt{t}(t)^2/t^2\right)$ is indeed analytic.

We recently~\cite{HathcockSnn} computed the free energy
Eq.~\ref{eq:2dfreeEnergy}, by perturbatively expanding Onsager's exact
free energy around the critical point. A key question is the radius of
convergence of the coordinate transformation $\wt t(t)$ and $f_a(t)$ to
the normal form. 
Unlike Taylor expansions about analytic points, the radius of convergence
of this normal-form analytic expansion about a singular point is not simply
the distance in the complex plane to the next-nearest singularity.
One might hope that physics would govern the convergence -- perhaps the distance
to zero temperature, infinite temperature, or the nearest other phase
transition. Indeed, in each of the expansions discussed below, the critical point is closest to zero temperature, with this distance determining the radius of convergence. 

\begin{figure}[t]
\centerline{\includegraphics[width=0.6\textwidth]{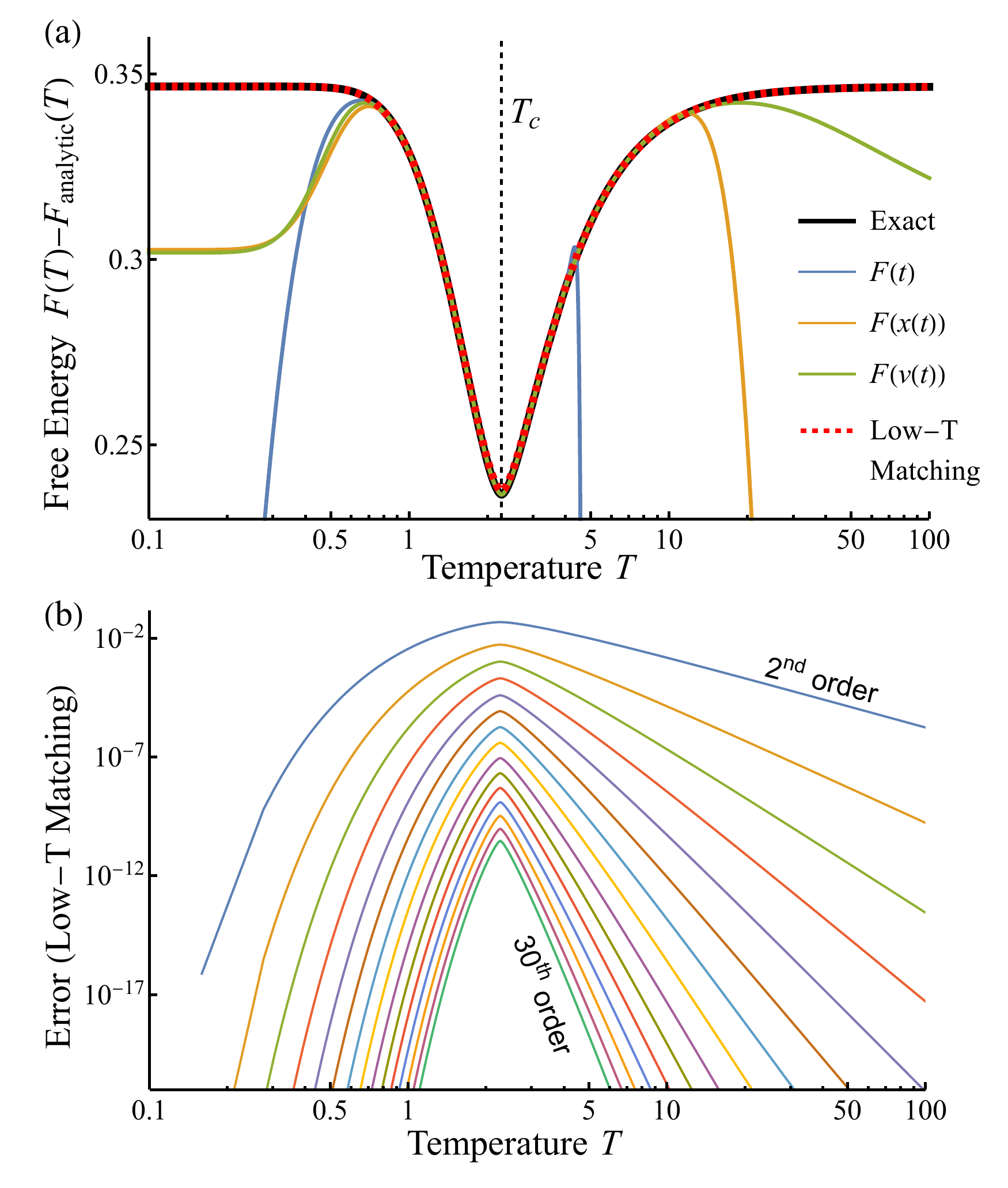}}
   \caption{{\bf Capturing the entire phase with analytic corrections, 2d Ising} (from~[\citen{HathcockSnn}]). 
   (a)~{\bf Radius of convergence depends on the expansion variable}. Expanding Onsager's exact free energy (black line) in $t$, $x$, and $v$ (colored solid lines) leads to increasing radii of convergence, with the convergence of the $v$-expansion estimated to cover the entire physical temperature range. Each expansion shown is $20^\text{th}$ order. Determining analytic corrections in $v$ by matching to the low-temperature expansion does even better, accurately reproducing the free energy across all temperatures even at low orders (red dashed line, $6^\text{th}$ order).
   (b)~{\bf Exponential convergence.} Adding analytic corrections to
   the universal scaling
   function at $T_c$ by fitting to the low-temperature expansion results in exponential convergence to the Onsager solution as we add more terms.
   \label{fig:HathcockExpConvergence}}
\end{figure}

Our investigations showed the importance of the choice of 
variable used to parameterize the distance to the critical point
(Fig.~\ref{fig:HathcockExpConvergence}(a)). Expanding $\wt t$ and $f_a$ in
temperature $t=(T-T_c)$, for example, converges in an estimated range
$-T_c < t < T_c$ (or $0<T<2T_c$). This is not the only natural choice, however. The
low temperature
expansions for the Ising model are expressed in powers of $X = \exp(-2/T)$.
$X$ is also a natural variable in that the zeros of the 2d
Ising partition function in the complex $X$ plane form a circle passing
through $X_c$, as Fisher explained~\cite{FisherCircleOfZeros}. 
Using the Onsager solution to expand $\wt t$ and $f_a$ in terms of $x = X-X_c$
yields a radius of convergence that extends all the way from zero temperature
to $X = 2 X_c$ (corresponding to $T \approx 4.7 \, T_c$), but fails to describe higher temperatures. 

Here we used special properties of the 2d Ising model to identify a new
variable
\begin{equation}\label{eq:vCoordinate}
    V = \frac{5-3 \sqrt{2} + X}{1+\sqrt{2} + X} \quad \quad  v = V-V_c,
\end{equation}
which allows our estimated radius of convergence to cover the
full physical temperature range, from zero to infinity
(Fig.~\ref{fig:HathcockExpConvergence}(a)). Our coordinate
$v$, unlike $t$ or $x$, respects the self-dual symmetry of the 2d Ising model.
Fisher's circle of zeros in $X$ (and thus $x$) breaks this self-dual symmetry.
The linear fractional transformation in Eq.~\ref{eq:vCoordinate} precisely
unwraps this circle of zeros into a straight line, extending the self-dual
symmetry to the complex plane. The circle of zeros becomes
the branch cut of the logarithm in the scaling function for the free energy
(Eq.~\ref{eq:2dfreeEnergy}).

Fig.~\ref{fig:HathcockExpConvergence}a compares the expansions $f(t)$, $f(x(t))$, and $f(v(t))$ to the exact free energy. In accordance with the discussion above, we see improved convergence for the $x$ and $v$ expansions. The expansion in $v$
numerically appears to have a radius of convergence that extends to the entire
physical range, zero to infinite temperature. Physics here does determine the
range of convergence of normal form theory.

The above results leave two questions: can we improve our approximation for the free energy further, such that it converges, even at low orders, across all temperatures? Furthermore, can we approximate the coordinate change and analytic background without knowledge of the exact solution or the nonlinear terms in the RG flows? To resolve these challenges, we again expand the free energy in $v$ using the form given in Eq.~\ref{eq:2dfreeEnergy}, but fix the expansion coefficients of $\wt t$ and $f_a$ by matching to low-temperature expansions of the free energy, instead of expanding the exact solution. Importantly, this approach requires minimal knowledge about the critical point (only the asymptotic scaling form), with most of the information coming from deep within the low-temperature phase. Because the matching guarantees the correct low temperature behavior and has the correct log-singularity at the critical point built into the expansion, we see uniform convergence across all temperatures. For example, by sixth order, the approximation differs from the true free energy by at most $0.5\%$ (red dashed line in Fig.~\ref{fig:HathcockExpConvergence}a) and we see exponential convergence as we add additional terms (Fig.~\ref{fig:HathcockExpConvergence}b). 

Our approach for extending critical scaling forms to the neighboring
phases should naturally generalize to unsolved statistical physics
models and experimental systems. For example, low-temperature expansions
are relatively easy to compute in all dimensions and for a variety of
systems and can be used to approximate the analytic corrections to
established critical scaling forms. Candidates for this method include
the 3d Ising model, where critical exponents are known to high precision
from conformal bootstrap and the 2d Ising model in a field, whose
scaling function is computed to high precision in the
section~\ref{sec:2dIsingInField} below. Finally, it is credible that the
liquid phase, long-known as being challenging because there is no `small
parameter', could be described as a perturbation of the liquid-gas
critical point, and described as the Ising critical point plus analytic
corrections (determined, for example, by matching to virial expansions).

\section{2d Ising critical point in a field}
\label{sec:2dIsingInField}

\begin{figure}
  \centerline{
    \includegraphics[width=\textwidth]{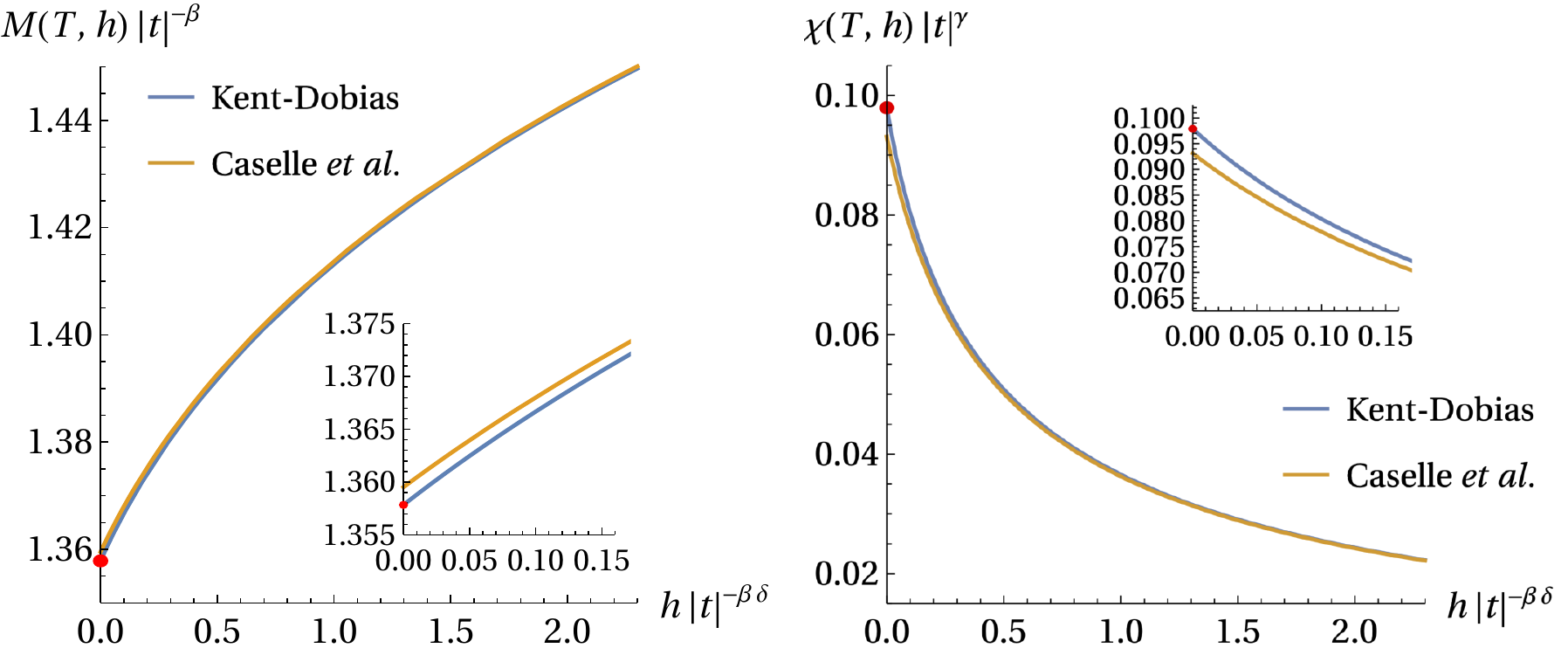}
  }
  \caption{
    {\bf Comparison between parametric approximations to the scaling functions with and without the appropriate singularities} (from~[\citen{KentDobiasSnn}]). Scaling functions for the magnetization and susceptibility plotted as functions of the scaling invariant $h|t|^{-\beta\delta}$. The errors in the blue curves (Kent-Dobias)
are estimated to be roughly $10^{-7}$. Caselle's earlier work~\cite{CaselleHPV01} 
that we build upon has significant errors near $h=0$ (red point);
these are due to the essential singularity at $h=0$ and $t<0$ that we address.
These discrepancies grow larger with higher derivatives of the free energy, as
shown in the right panel of Fig.~\ref{fig:JaronExpConv}.
}
  \label{fig:JaronVsCaselle}
\end{figure}

Ever since Onsager solved the zero-field 2d Ising model, people have searched
for a high-precision approximate solution for the 3d Ising model,
and occasionally also for the 2d Ising
model in a field. The most successful of these attempts make use of parametric
coordinates to interpolate behavior in temperature and field around the critical
point in such a way that the singularity is naturally incorporated.
\cite{GuidaZj97, CaselleHPV01} These coordinates, first introduced many decades
ago, \cite{Schofield_1969_Parametric} are polar-like, with a radius-like
component $R$ that controls the proximity to the critical point, and an
angle-like coordinate $\theta$ that rotates around the temperature--field
plane. When taken to parameterize the temperature--\emph{magnetization} plane,
such coordinates can be periodic in $\theta$ like true polar coordinates,
something once studied by Michael and collaborators to describe the coexistence
region. \cite{Fisher_1999_Trigonometric} However, when parameterizing the
temperature--\emph{field} plane, there is an inevitable cut resulting from the
discontinuity between the up and down states at low temperatures, and $\theta$
is not a periodic coordinate but ends at the abrupt transition at some
$\theta_0$.

The power of adopting these coordinates is that, as a function of $\theta$, the
critical singularity does not appear, and one can reasonably well-approximate
the scaling function in this coordinate as a simple analytic function. Using
this principle, Caselle \emph{et al.}\ did a remarkably careful job of creating
a function that satisfied lots of known properties and measured values (in
yellow of Fig.~\ref{fig:JaronVsCaselle}). Their method was doomed to slow convergence at small external fields,
though, because while the singularity of the critical point was indeed removed
by the coordinate change, other more subtle singularities in the free energy
remain, including a key essential singularity as one crosses the $h=0$ line
for $t<0$.
Fig.~\ref{fig:JaronVsCaselle} shows a comparison between the approximate form of
Caselle \emph{et al.}\ and that of our work including this singularity.

\begin{figure}
\centerline{\includegraphics[width=0.5\textwidth]{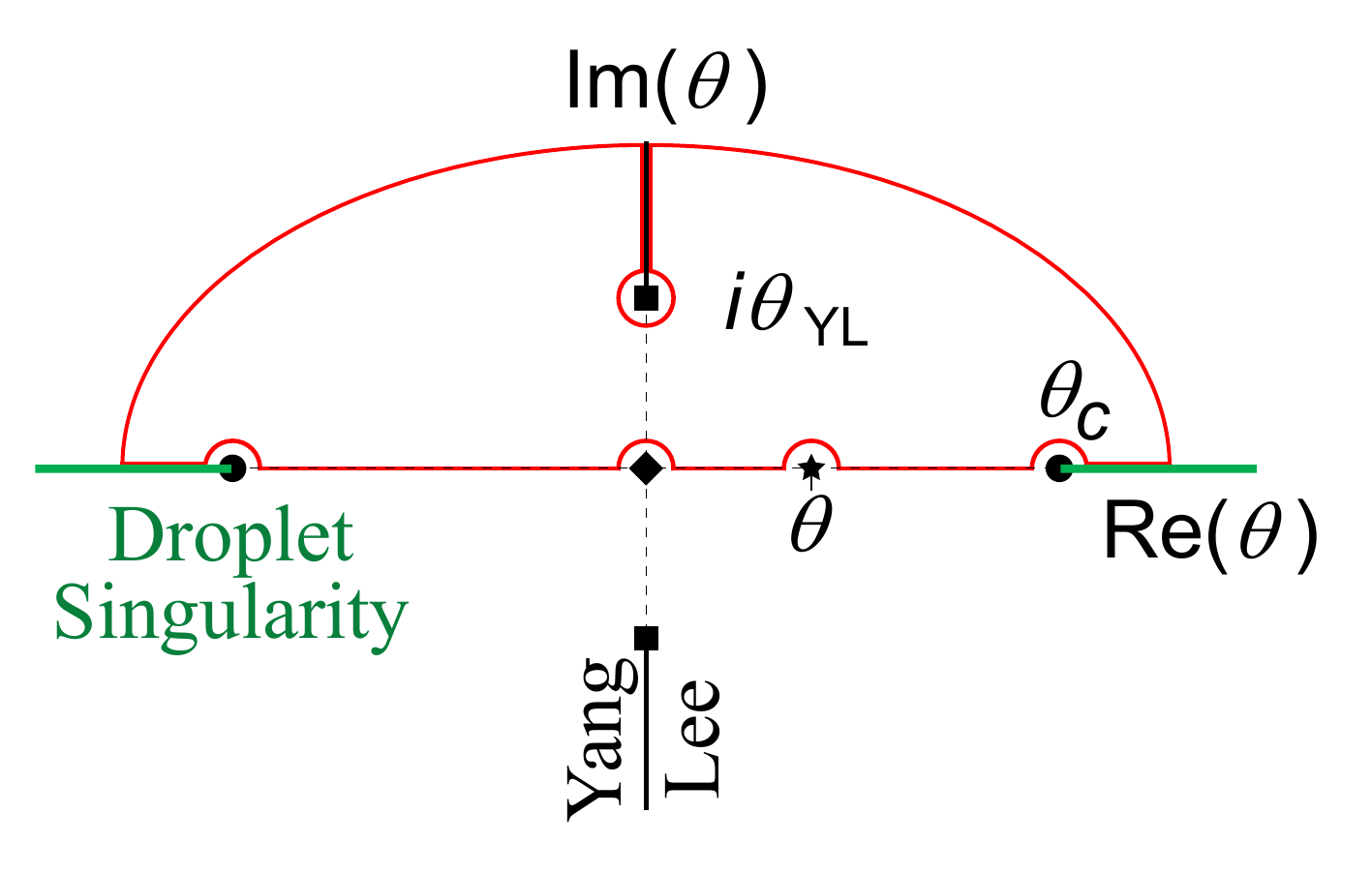}}
   \caption{{\bf Contour in the complex $\theta$ plane} to generate a free energy
   scaling form with the correct singularities (from~[\citen{KentDobiasSnn}]). 
   \label{fig:JaronContour}}
\end{figure}

In order to rigorously realize the idea of a free energy scaling function
broken into singular and analytic pieces, two singularities in the scaling
function must be accounted for in addition to the critical one.
One is ancient and well known to the experts: the Yang--Lee edge singularity,
for $T>T_c$ in the complex plane.\cite{Yang_1952_Statistical,
Lee_1952_Statistical} The other has only recently been realized to be
relevant, but has a nice physical picture. As one crosses $h=0$, the
equilibrium magnetization jumps. But if you cross just a little bit, the
up-spin magnetized state is metastable, lasting for a good while until a bubble
of down spins forms and grows to flip the system. (Think of supersaturated
vapor -- 101\% humidity -- and the nucleation of raindrops.)
Here the surface tension cost between the up and down regions for small droplets
is bigger than the bulk free energy gain of aligning the interior with the
external field. The energy barrier $B$ to reach large sizes diverges
diverges as $h$ vanishes, $B/k_B T = c/h$ for some constant $c$.
Just like in quantum mechanics, where the lifetime
of a state gives the energy an imaginary part, the free energy becomes complex
for $h<0$ with an essential singularity $\mathrm{Im}{f} \sim e^{(c/h)}$. One can use a
Kramers--Kronig transform to see that the real part also has an essential
singularity, influencing $f$ for $h>0$ as well.

\begin{figure}
  \centerline{
    \includegraphics[width=\textwidth]{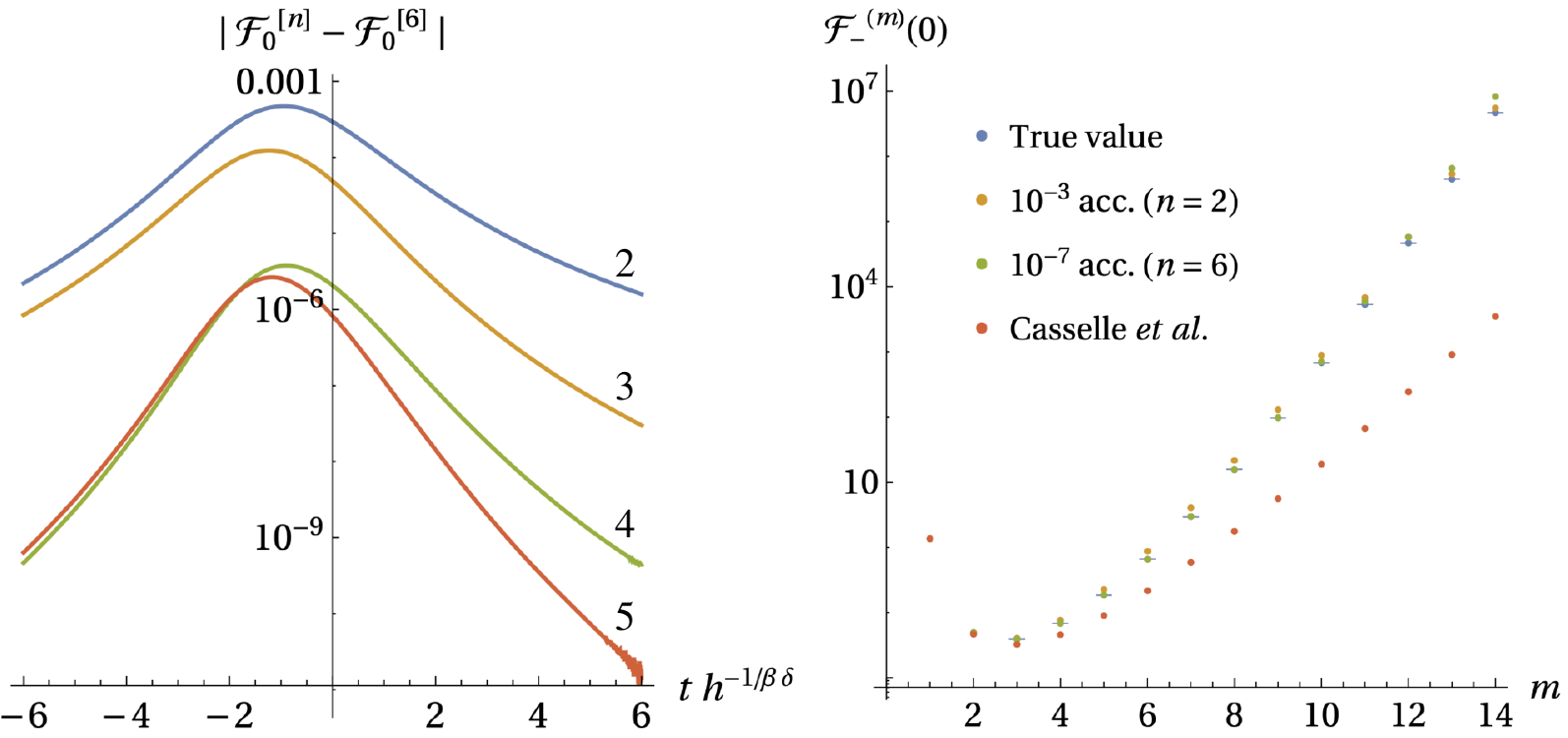}
  }
  \caption{
    {\bf Convergence of the universal scaling function} (from~[\citen{KentDobiasSnn}]).
    Left: The difference between the $n$th order approximation for the free energy scaling function in a field and the 6th order approximation as a function of the scaling invariant $th^{-1/\beta\delta}$. The 5th and 6th order approximations differ by at most $2\times10^{-6}$. Higher derivatives behave similarly. Right: The Taylor series coefficients $\mathcal F_-^{(m)}(0)$ of the free energy scaling function at the abrupt transition line (the location of the essential singularity) as a function of derivative $m$. By incorporating the essential singularity into the scaling function explicitly, the approximate forms have zero radius of convergence, matching numeric measurements of these coefficients.
  } \label{fig:JaronExpConv}
\end{figure}

We can incorporate these two singularities into the universal scaling function
by first generating a `simplest' functional form that has the correct
singularities, and then changing variables $\theta \to \wt{\theta} = g(\theta)$
and adding an overall analytic function $\mathcal{F}_a(\theta) = G(\theta)$,
in a precise analogy with the analytic corrections we introduce to match
the universal scaling function to describe the surrounding phases 
(sec.~\ref{sec:AnalyticCorrections}).

The `simplest' form is taken from writing the most
singular part of the imaginary part of the free energy, and then taking
advantage of a Kramers--Kronig like relation to find the corresponding real
part. The requisite contour integral in the complex-$\theta$ plane is shown in Fig.~\ref{fig:JaronContour}.

Though complicated, this procedure pays dividends. By incorporating these singularities and matching the power series terms in $g(\theta)$ and $G(\theta)$ (analogous to $\wt t(t)$ and $f_a(t)$ in section~\ref{sec:AnalyticCorrections}), we are able to achieve exponential convergence of the scaling function to exactly known values at $t=0$, and also achieve exponential convergence of derivatives. The left-hand side of Fig.~\ref{fig:JaronExpConv} shows this convergence in the free energy itself as a function of the scaling invariant $th^{-\beta\delta}$ by 
subtracting our best converged approximation $\mathcal{F}^{[6]}$ from the
lower-order approximations $\mathcal{F}^{[n]}(\theta)$ for $n\in \{2,3,4,5\}$.
These provide evidence that our seven-digit convergence at $t=0$ extends to 
the whole scaling function.
On the right of this plot, we can see the origin of this good behavior: the series expansion for the free energy around the abrupt transition point has zero radius of convergence, but this is captured naturally by our approximate scaling form.

\section{Interpolating scaling functions between dimensions}
\label{sec:NPFRG}

\begin{figure}
\centerline{\includegraphics[width=0.5\textwidth]{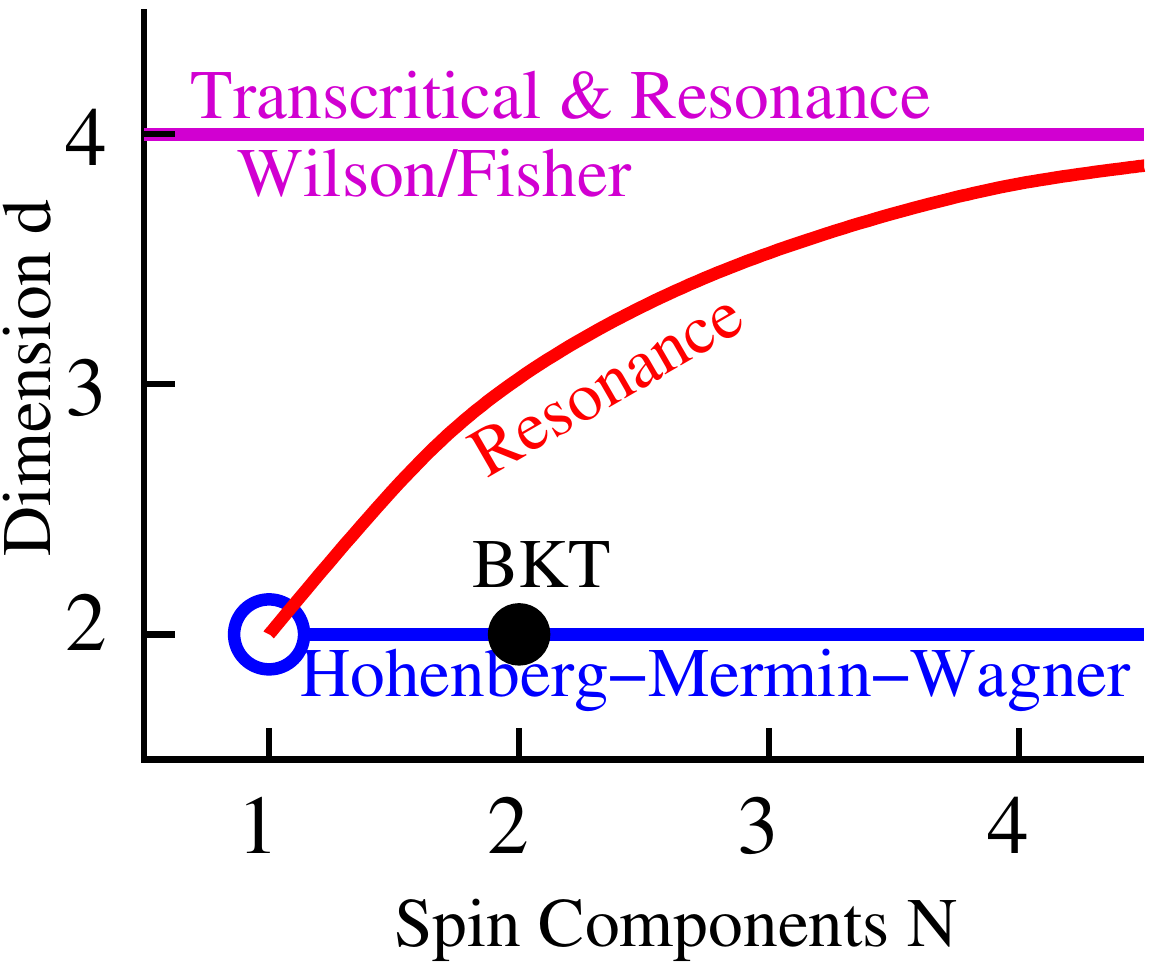}}
   \caption{{\bf Interpolating between dimensions.} A schematic
   showing the dimensions and spin components where traditional power-law
   scaling will break down. This leads to logarithms and exponentials replacing
   power laws, and rich but complicated invariant scaling combinations as
   arguments for the universal scaling functions~\cite{RajuCHKLRS19}. 
   On the other hand, these complexities also promise to inform and accelerate
   convergence of the universal scaling functions for dimensions in between.
   \label{fig:WFPhaseDiagram}}
\end{figure}

Figure~\ref{fig:WFPhaseDiagram} illustrates the traditional thermodynamic
critical points. Over five decades, these have been extensively explored
using $\epsilon$-expansions, $1/N$ expansions, cluster expansions, etc.
The universal critical exponents are known essentially exactly from conformal
bootstrap methods.%
  \footnote{In the same spirit, Kent-Dobias' solution~\cite{KentDobiasSnn}
  for the scaling function in 2d in section~\ref{sec:2dIsingInField} 
  is also essentially exact: both are well-defined algorithms that generate
  an exponentially converging approximation in a form useful for applications.
  After all, this is what we call the `exact' solution for $\sin(x)$.}
Can we do the same for the universal scaling functions
as functions of dimension $d$ and the number of spin components $N$?
And for (say) the random-field Ising model~\cite{RFIMNPFRG20}, or
turbulence~\cite{TurbulenceNPFRG21}?

Interpolation between dimensions could have several benefits. First, 
much is known analytically in two and four dimensions about the universal
scaling functions that is only known numerically in three. Second, there
are important features in scaling functions and their corrections to scaling
that are a clear foreshadowing of properties in other dimensions. So the 
leading correction to scaling in 3d is the echo of the marginal variable
of the Wilson-Fisher transcritical bifurcation in 4d. And, in a more 
dramatic example, the universal scaling function for the avalanche size
distribution in the 3d random-field Ising model has a striking feature 
(it grows by a factor of ten before cutting off
exponentially~\cite{PerkovicDS99}), which
is clearly related to the unusual scaling of the avalanche sizes in two
dimensions~\cite{HaydenRS19} (Fig.~\ref{fig:AvalancheSizeCollapse}).

In ordinary differential equations, normal form theory provides not only
the behavior at the transition, but also an unfolding of the behavior near
the bifurcation (up to a smooth coordinate transformation). Here we may
expect to use this, not just to describe the phases near the critical points,
but (following the lead of Wilson and Fisher) to describe the universal
scaling functions as they evolve between dimensions. 

How would we interpolate scaling functions, when even the arguments of the
scaling functions vary with dimension (as in the $W$-functions we needed
in section~\ref{sec:NormalForms} for the 4d Ising model)? Consider taking
the 4d flow equations~\ref{eq:FlowEquationIsing4D} about the mean-field
fixed-point, and keeping the unremovable nonlinear terms to describe
nonlinear flows about the mean-field fixed point in $d$ dimensions:
\begin{equation}
  \label{eq:FlowEquationIsingAllD}
  d{\wt{f}}/d{\ell} = d \wt{f} - \wt{t}^2, ~~~ 
  d{\wt{t}}/d{\ell} = 2\wt{t} - A \wt{u} \wt{t}, ~~~
  d{\wt{u}}/d{\ell}  =(4-d) \wt{u} - \wt{u}^2 + D \wt{u}^3, ~~~ 
\end{equation}
Things to note. 1)~The normal-form procedure of removing the nonlinear terms
allows us to set the nonlinear terms in any way we wish. 2)~We know the
Wilson-Fisher fixed point for Eqs.~\ref{eq:FlowEquationIsingAllD} at 
$(\wt{t}^*,\wt{u}^*)$ must have a linearization whose eigenvalues $1/\nu$
and $\omega$ that, as a function of dimension, matches the conformal bootstrap
values (or the exact values in 2d). This constrains the two nonlinear
terms $A(d)$ and $D(d)$. Serendipitously, the $\wt{t}^2$ resonance term 
in the 4d free energy flow is also needed to get the logarithmic
specific heat in 2d. A high-precision universal scaling function for
the free energy using the nonlinear scaling variables $\wt{f}$,
$\wt{t}$ and $\wt{u}$ in
Eq.~\ref{eq:FlowEquationIsingAllD} would describe the universal crossover
scaling from mean field to short-range magnetization
in all dimensions between two and four. One could then linearize that
scaling function about the 3d Wilson-Fisher fixed point to extract the
traditional scaling functions.

How would one calculate such scaling functions? Here serendipity strikes in
recent advances in non-perturbative functional renormalization-group (NPFRG)
calculations~\cite{NPFRG21}, whose critical
exponents~\cite{CriticalExponentsPrecisionNPFRG20} are almost competitive
with those of conformal bootstrap, but which in the process also compute
a functional form for the coarse-grained free energy as a function of
magnetization.
 
Matthew Tissier and colleagues at the Jussieu campus of the Sorbonne and the
first author
have been exploring how to calculate high-precision universal scaling functions
by combining the strengths of the normal-form theory with the systematically
improvable scaling functions of NPFRG. In initial work, we have found that
the NPFRG for the 4d Ising model should indeed have a scaling function
whose arguments are of the exotic form in Eq.~\ref{eq:InvariantCombinations4D}.
We are now learning to solve the partial differential equations for 
$\wt{t}(t,u)$ $\wt{u}(t,u)$ and $\wt{f}(t,u)$ in 3d to extract the universal
scaling functions from the simplest of NPFRG models. We hope then to use
the technology in section~\ref{sec:2dIsingInField} to demonstrate how 
to tabulate universal scaling functions in a form convenient for 
experimentalists and simulators.

The NPFRG method has had striking success in high-precision calculations
of equilibrium thermodynamic systems, disordered thermodynamic and 
avalanche models, fully developed turbulence, quantum many-body systems, 
and in QCD and electroweak models~\cite{NPFRG21}. Our hope and vision
is to inspire our colleagues to gather and tabulate their results into
universal scaling functions that can describe the behaviors both near
and far from points of emergent scale invariance, in a way accessible to
engineers, biologists, and social scientists studying complex systems.

\section{Prospects and challenges for future work}
\label{sec:ProspectsForFuture}

We have summarized what we believe to be promising indications that 
we can indeed generate usable, high-precision universal scaling functions in
a systematic way, and can extend them systematically into high-precision
descriptions of the surrounding phases. If this continues to work, one
imagines our community will be able to provide quantitative theories
of liquids, materials plasticity and failure, fluctuations in biological
systems, and potentially turbulence and glass behavior. In all these
cases, we must mesh our understanding of the universal, emergent
critical behavior and our system-dependent knowledge of properties away 
from criticality to describe the entire phases. Will this work in practice? 

To show that our method works to high precision, we chose to use a problem
where we knew a great deal about the answer: the 2d Ising model.
The reader should legitimately be
skeptical that this is the easiest case. In particular, the 2d Ising mode
is special in that it (1)~is self-dual, (2)~is analytically solvable in
zero field, and (3)~has no singular corrections to scaling.
Let us discuss causes for concern that our methods will become general tools,
and reasons for optimism.

(1) {\em The 2d Ising model is self-dual.} 
In section~\ref{sec:AnalyticCorrections}, 
we found exponential convergence of the Ising free energy $f(t)$ in zero field
using a power series of the Onsager solution about the critical point.
To do so, we expanded in a variable $v(t)$ that was fine-tuned to the
self-dual and complex analytic properties of the 2d Ising model. We then
used the low-temperature expansion to avoid any use of Onsager's solution.
Will we need to find the `right' variable in 3d to proceed? 
It turns out that matching to properties outside the radius of convergence
stabilizes the expansion. In preliminary work,
matching both high and low temperature series using the standard
variable $x = X-X_c = e^{-2/T}-e^{-2/T_c}$ in 2d (ignoring duality), 
we found exponential convergence in the entire temperature
region~\cite{HathcockSnn}. 

(2) {\em The 2d Ising model in zero field has an exact solution},
and also has an exact solution at $T_c$ as a function of field.%
  \footnote{It may seem in retrospect clear that the essential
  singularity at $h=0$ would frustrate a search for an exact solution
  in a field below $T_c$.}
But will the external field prevent us from finding an exponentially
converging solution using analytic corrections to scaling
$\wt{t}(t,h)$, $\wt{h}(t,h)$, and $f_a(t,h)$, by matching the 
2d universal scaling function from
sec.~\ref{sec:2dIsingInField} to the low $T$/high~$h$ and high $T$/low~$h$
cluster expansions, as we did for zero field in 
sec.~\ref{sec:AnalyticCorrections}?

We have an exact solution for the magnetization -- the first derivative
with respect to field. This has allowed us to determine $\wt{t}$ and $\wt{h}$
to linear order in $h$.
The susceptibility $\chi(T)$ at zero field is not known exactly, but a 
remarkable amount is known~\cite{ChanGNP11}:
there are formal expansions strongly indicating subdominant logarithmic
corrections to scaling, and a possible
essential boundary in the complex plane at the Fisher circle of zeros 
in the complex temperature plane~\cite{Nickel_1999_On,orrick2001susceptibility,
Assis_2017_Analyticity,ChanGNP11}. Log corrections
to scaling are known to arise from resonances between relevant and irrelevant
eigenvalues (sec.~\ref{sec:NormalForms}); the observed logs could arise from
irrelevant variables whose resonances contribute only in $O(h^2)$ to the 
free energy. It appears that we can use these results~\cite{ChanGNP11}
not only to determine $\wt{t}$ and $\wt{h}$ to quadratic order in $h$,
but also to extract information about irrelevant variables and their singular
corrections to scaling in 2d (see below).

The universal scaling function in 3d presumably should succumb to the same
kind of systematic approximation that we used in
sec.~\ref{sec:2dIsingInField} for 2d. All the ingredients appear to be available.%
  \footnote{And, of course, high-precision NPFRG calculations of the
  critical exponents~\cite{CriticalExponentsPrecisionNPFRG20}
  implicitly have calculated the universal scaling function.}
The Yang-Lee singularity in 3d~\cite{YangLeeNPFRG20} has yielded
to a high-precision NPFRG calculation. The essential singularity calculation
depends only on the value of the surface tension near the critical point.
In 2d, we used estimates of derivatives of the universal
scaling function near $h=0$ for $T>T_c$ and $T<T_c$. The latter
(shown in Fig.~\ref{fig:JaronExpConv}b) are challenging to calculate
using NPFRG methods (perhaps because of the essential singularity), but
the former are now available~\cite{DePolsi_2021_Precision}.
Indeed, these authors
used their estimates and the traditional Schofield coordinates to estimate
the scaling function; adding the proper droplet singularity and 
Yang-Lee singularities could be
straightforward and hopefully will yield convincing exponential convergence.

Can we expect to extend a high-precision 3d Ising scaling funciton to 
describe the in-field behavior in the entire surrounding phases?
Using analytic corrections to extend the critical singularity
could break down at the roughening transition.%
  \footnote{While one might argue
  that the behavior of an interface between spin up and spin down regions 
  should not affect the bulk free energy, the (incredibly subtle)
  essential singularity as one crosses $h=0$ (sec.~\ref{sec:2dIsingInField})
  depends on the surface energy of the critical droplet, which will itself
  have an essential singularity as a function of temperature as it develops
  facets at the roughening temperature.}
However, this subtle transition should in principle also impede the use
of Dlog Pad{\'e} methods using the low-temperature expansion of the magnetization
to describe the critical-point behavior, which works well in practice.
In any case, the roughening transition does not arise
in isotropic systems, so these concerns will not interfere with finally
extracting a quantitative theory of liquids by adiabatic continuation of our
understanding of the 3d Ising critical point.

(3) {\em The Onsager solution exhibits no singular corrections to scaling.}
First, this statement is more subtle than it seems. Onsager's solution
has a pure logarithmic singularity, but conceivably there could be irrelevant
eigenvalues with integer exponents that would lead to analytic terms in the
free energy and magnetization. Also, there are irrelevant anisotropies in the
correlation functions and the susceptibilities due to the square lattice, 
indeed with integer exponents.  Even more subtle,
these anisotropies are singular corrections to scaling in Wilson's
RG formulation, but are clearly present at the fixed point of the 
real-space RG. Any coarse-graining procedure maintaining a square lattice
will have a short-range square anisotropy.

Experts will
remember that some of the irrelevant operators in the RG are {\em redundant},
and these redundant operators often involve flows between different fixed
points in the same universality class.%
  \footnote{We have recently looked at this question~\cite{RajuSnn} in
  the context of the period doubling onset of chaos in iterated maps $f(x)$.
  About half of the irrelevant eigenvalues are negative powers of the 
  relevant critical exponent $\alpha$, and are redundant: one can set
  the amplitudes of their correction to scaling to zero using a coordinate
  change $x = \phi(y)$, so instead of $\alpha f(f(x/\alpha))$ one considers the 
  renormalization group $\alpha \phi(f(f(\phi^{-1}(x)/\alpha))$. We called
  these {\em gauge irrelevant}, as they depend on the choice of coordinate.
  A similar gauge invariance arises for the quasiperiodic onset of
  chaos~\cite{RandOSS82,OstlundRSS83} that we studied when the first 
  author was a postdoc.} 
The idea of redundant variables
had been first explored by Wegner~\cite{wegner1974some}, who pointed out 
that a RG transformation would always have certain operators that are 
redundant; they correspond to infinitesimal redefinitions of the field, and
do not contribute to the singular part of the free energy.%
  \footnote{How does one think of these redundant operators in the discrete
  theory? Murthy and Shankar had found a clever way to think about redundant 
  operators for the discrete Ising model~\cite{murthy1985redundant}. They 
  essentially used a version of Kadanoff's RG projection operator without 
  reducing the degrees of freedom. They found evidence that these redundant 
  operators had observable consequences in simulations of the 2d Ising 
  model~\cite{shankar1985clear}. Evidence for redundant operators had 
  already been found in the 3d Ising model~\cite{pawley1984monte}.}

The first author's first graduate student, Mohit Randeria, worked for Michael
Fisher before shifting to my group. The two of them wrote a comment 
on a paper by Swendsen~\cite{swendsen1984optimization} claimed that one could
modify the RG transformation to move the fixed point anywhere on the 
critical surface. Randeria and Fisher asserted that this could not in 
general be correct, because the amplitudes of the singular corrections to 
scaling are zero at the fixed points. To quote them, 
\begin{quote}
``One fixed point
may be mapped into another by a change of redundant variables and two RGs, 
say $\mathcal{R}$ and $\widehat{\mathcal{R}}$, may produce {\em formally}
different fixed points by this mechanism. Nevertheless, a general point on
a critical manifold {\em cannot} be transformed into a fixed point nor
{\em vice versa}.''
\end{quote}

In the context of the 2d Ising model, there seem to be no singular corrections in the exact solution. Barma and Fisher had predicted that there should be  logarithmic corrections to scaling coming from irrelevant variables~\cite{barma1984corrections}. They had also found evidence for an irrelevant operator with exponent $-4/3$~\cite{barma1985two}. Conformal field theory predicts a large number of irrelevant variables which come from descendant operators but does not predict the $-4/3$ exponent. It has been conjectured in the past that all of these descendent operators are redundant~\cite{blote1988corrections}. This may explain why they do not lead to any logarithmic corrections to scaling (which otherwise are generically expected~\cite{RajuPhD}). We would consider all of these descendant operators as constituting `gauge' corrections to scaling, which can be removed by an appropriate coordinate choice, whereas the $-4/3$ exponent would be a genuine singular correction to scaling. 

In sec.~\ref{sec:AnalyticCorrections} we succeeded in solving for the free energy of the square-lattice Ising model at all temperatures by applying a normal-form change of variables to the universal scaling function for the critical point. Is that success due to the lack of (non-redundant) singular corrections
to scaling for this particular universality class?%
  \footnote{It is believed that the powers of logarithms in the 2d Ising
  susceptibility vanish in isotropic systems.
  Since all irrelevant operators in 2d have rational eigenvalues,
  and the redundant ones cannot generate resonant logarithms, one suspects
  that all the isotropic corrections to scaling are redundant.}
It is clear that in 3d we shall need to incorporate the (genuine) singular
corrections to scaling in order to accurately describe the behavior away from
the fixed point. Indeed, these contributions are subdominant near the 
fixed point, meaning conversely that they grow faster as one leaves the fixed point 
than do the relevant operators. Thus we will need a universal scaling function
that depends not only on $\wt{t}$ and $\wt{h}$, but on potentially an
infinite family of irrelevant variables $\wt{u}$. Can we nonetheless hope for exponential
accuracy?

The NPFRG methods mentioned in sec.~\ref{sec:NPFRG}, as it happens,
precisely implement a flow with many 
irrelevant eigendirections at the fixed point. Applying our normal form
analysis, we have been able to linearize this flow in the simplest case,
allowing us to extract a functional form for the leading correction
subdominant as $\widetilde{t}^\omega$. It seems likely that one could generate
exponential convergence for all properties of the phases by combining
increasingly sophisticated NPFRG calculations, adding
more irrelevant operators to our scaling functions, and matching to 
higher orders of low and high temperature expansions (or virial expansions, 
$1/N$ expansions \dots). 

Finally, there are intriguing hints of deep connections between normal form
theory, redundant operators, and the construction of universal scaling functions.

We noted in sec.~\ref{sec:2dIsingInField} that the procedure we used to generate
the universal scaling function recapitulated exactly the same steps as
the procedure we used in sec.~\ref{sec:AnalyticCorrections} to generate
the Onsager solution from the critical point. In both cases, we knew the 
singularities of the functions involved, and fit parameters in two functions
(an analytic change of variables and an additive analytic background) to 
match known properties measured separately. If this systematic procedure can
be generally used as a best practice, rapid progress could be made.

There is also a striking analogy between redundant operators (removing 
some singular corrections to scaling by reparameterizing the space of
predictions)
and our normal form theory (removing analytic corrections to scaling by 
reparameterizing the control parameters). Can we simultaneously remove both
with a joint transformation? How are they related?

In thermodynamics, whether you write the free energies as a function of the
external field and temperature or as a function of magnetization and temperature
is considered a matter of choice. A Legendre transformation from control
parameters $(f,t,h)$ to $(u,t,m)$ swaps the control parameter $h$ with the 
prediction $m$: does it swap normal-form transformations to redundant ones?
(Indeed, the NPFRG methods coarse-grain the fixed-magnetization Gibbs free
energy $u(t,m)$, rather than Wilson's free energy at fixed field.)
When we use our normal form theory to remove nonlinear terms in the RG
flows by changing $h$ to $\wt{h}(t,h)$, we change the predicted magnetization.
Is that change a redundant one?

Hankey and Stanley~\cite{hankey1972systematic} showed that if you
Legendre transform the generalized homogeneous functions we use for
hyperbolic fixed points, you get another homogeneous function. Later,
others, studying finite-size scaling of the Ising
model~\cite{desai1988finite, kastner2000microcanonical} in the microcanonical
ensemble, predicted that it would be possible to speak of the entropy as a
scaling function of the energy instead of the free energy as
a function of the temperature.

Our preliminary work on this question ran into a difficulty, generalizing
these ideas to the nonlinear normal forms necessary in the upper and lower
critical dimensions.
We have found that it is possible to Legendre transform RG flows at the
linear level, but nonlinear flows generically lead to non-analyticities
in the RG flow~\cite{RajuPhD, ClementPhD}.%
  \footnote{As noted in the footnote on page~\pageref{foot:NonlinearTermHZero},
  even in four dimensions the RG flow of the magnetization is linearizable.
  Presumably this means that implementing an RG using flows in 
  $(f,t,h)$ and $(u,t,m)$ will be possible
  even in four dimensions, while we find $(f,t,h)$ to $(S,E,h)$ leads
  to non-analytic RG flows.}
Whether this clarifies or confuses the correspondence between redundant
and normal-form changes of variables needs to be explored.

As a long-time colleague of Michael Fisher, the lead author hopes that Fisher
would have enjoyed this deep plunge into the complex history of the field 
and our ambitions for the future. He also hopes that Fisher's friends,
colleagues, and collaborators who are contributing to this book, and of course
our colleagues of the future who are reading it, will find this chapter 
more illuminating than obscure, and more useful than misleading or misguided.

\section*{Acknowledgments}
JPS would like to thank Matthieu Tissier for his insights and patience explaining
the NPFRG methods, and for continuing collaborations in this work, Colin
Clement for careful early work on the 2d Ising model, and Jacques Perk for
helpful correspondence. JPS
benefited by funding from NSF DMR-1719490 and CNRS (the French National Center
for Scientific Research). DH was partially supported by an NSF Graduate Research Fellowship Grant No. DGE-2139899. JK-D acknowledges Simons Foundation Grant No. 454943. AR acknowledges support from the Department of Atomic Energy, India under project No. RTI4006, and the Simons Foundation Grant No. 287975.
 
\bibliographystyle{ws-rv-van}
\bibliography{SethnaRecs,fisher,Memorial}

\end{document}